# Spectra of Earth-like Exoplanets with Different Rotation Periods

S. I. Ipatov[a,] * and J. Y-K. Cho[b]

[a] *Vernadsky Institute of Geochemistry and Analytical Chemistry, Russian Academy of Sciences, Moscow, 119991 Russia*
[b] *Department of Physics, Brandeis University, Waltham, MA 02453 USA*
*e-mail: siipatov@hotmail.com

**Abstract**—At present, planets like the Earth have been found near other stars. We investigate the spectra of Earth-like planets but with different axial rotation periods. Using the general circulation model of the atmosphere called the Community Climate Model (CCM3) and considering the atmospheric circulation lasting for two years, we calculated the radiation spectra of the Earth and the exo-Earth rotating with periods of 1 and 100 days, respectively. The radiation spectra of the atmospheres were calculated with the SBDART code. We analyzed the spectrum of upward radiation at altitudes of 1 and 11 km in wavelength ranges of 1 to 18 μm and 0.3 to 1 μm. The following common features were obtained for the Earth and the exo-Earth: (1) the planets exhibit a wide absorption band of $CO_2$ around 14 μm; (2) the radiation spectra at different locations near the equator show no significant differences (however, for some regions, e.g., near the poles, there can be considerable differences in the spectra); and (3) if the spectrum is integrated over the entire disk of the Earth/exo-Earth, the difference in the spectral signal obtained in observations from different directions becomes substantially lower than the difference between the results of observations of individual regions of the planets; however, the difference in the integrated signal of the spectrum for the Earth and the exo-Earth is noticeable (for example, this difference is noticeable for the spectrum obtained at an altitude of 11 km, when observing the South and North Poles; though, the difference is small, if one observes the whole disk from different equatorial directions). The differences in the spectra of exoplanets, which differ from the Earth only in axial rotation period, are comparable to the differences associated with changes in the angle of viewing the planet. Consequently, if the observation angle is not known, the analysis of the spectrum of the planet cannot be used to determine its axial rotation period. The maximal differences in the spectra of Earth-like exoplanets were obtained for wavelengths of about 5–10 and 13–16 μm. By analyzing the spectrum at wavelengths around 9.4–10 μm, we can determine whether the atmosphere of the exoplanet contains ozone or not. In the diagrams for the upward radiation at an altitude of 11 km, there is no local minimum at wavelengths of 9.4–10 μm if ozone is absent; and, when the models contain ozone, this minimum is present. Since ozone is essential for life, the 9.4–10 μm band may be important for future observations of Earth-like exoplanets.

## INTRODUCTION

Planets like the Earth have currently been found near other stars (Ananyeva et al., 2022; Marov and Shevchenko, 2020, 2022; Covone et al., 2021; Godolt et al., 2019; Ormel et al., 2017; Ribas et al., 2016). Some exoplanets (e.g., Kepler-62f, Kepler-186f, Kepler-442b, Kepler-1649c, Proxima Centauri b, and TRAPPIST-1e) are thought to have a chance of being habitable. To estimate the possibility of detecting signs of life in Earth-like planets, we study the spectra of these planets, but assuming they have different axial rotation periods. These studies provide a better understanding of what our planet or a similar exoplanet with a large axial rotation period might look like through a telescope.

The problem of the Earth as an exoplanet and observations of the Earth from space have been discussed in a number of papers (e.g., Olson et al., 2018; Robinson and Reinhard, 2019; Mettler et al., 2020, 2023, 2024). Additional references to publications on this subject can be found in the above listed papers. When searching for life on planets, Olson et al. (2018) suggested that one should pay attention to seasonal changes in the composition of the atmosphere ($CO_2$, $CH_4$, $O_2$, and $O_3$). Robinson and Reinhard (2019), in particular, presented the spectra of the Earth obtained in observations carried out with the *Galileo*, EPOXI, and LCROSSS spacecraft, with the Atmospheric Infrared Sounder (AIRS) onboard the *Aqua* satellite (supported by NASA) orbiting the Earth, and with the Thermal Emission Spectrometer (TES) onboard the Mars Global Surveyor (MGS). In the mentioned paper, it was noted that the climate of the Earth and its atmosphere were changing with time.

Mettler et al. (2020) used the data of observations of the Earth collected with the Moderate Imaging Spectroradiometer (MODIS) onboard the *Aqua* satellite. A complete dataset covers 15 years of observations of thermal radiation in the 3.66–14.40 μm range for five different locations on the Earth (the Amazon rainforest, Antarctica, the Arctic, the Indian Ocean, and the Sahara Desert). Mettler et al. (2020) showed that: (1) the geometry of observations plays an important role in the analysis of thermal radiation data, since the spectrum of the Earth varies by a factor of three or more depending on the dominant type of the underlying surface; (2) the typically strong absorption bands of $CO_2$ (15 μm) and $O_3$ (9.65 μm) are much less pronounced and partially absent in the data from polar regions, which suggests that, in these cases, it may be a challenge to correctly estimate the abundance levels for these molecules.

Mettler et al. (2023) analyzed infrared spectra obtained from the AIRS satellite during a four-year period (2016–2019). They showed the spectra between 3.75 and 15.4 μm for four directions of observations of the Earth (from the North and South Poles and from the equator with a center in the Pacific Ocean and in Africa). The radiation was slightly larger in summer than in winter. Mettler et al. (2023) noted that, at the peak wavelength of the Earth's radiation at ~10.2 μm, the view of the Earth from the North Pole and the view from the equator centered in Africa showed an annual variability of 33 and 22%, respectively. On the other hand, the observations of areas with a high fraction of water surface (e.g., the view from the South Pole and the view from the equator centered on the Pacific Ocean) show less annual variability due to the large thermal inertia of the oceans. Mettler et al. (2024) discussed the explorations of the Earth in the mid-infrared range with the Large Interferometer For Exoplanets (LIFE) designed for a future space mission. In their opinion, studies aimed at detecting the possibility of life on exoplanets should mainly focus on static evidence for life (e.g., the coexistence of $O_2$ and $CH_4$). Their main findings indicate that one should look for a habitable planet with significant abundances of $CO_2$, $H_2O$, $O_3$, and $CH_4$.

The James Webb Space Telescope (JWST) allows infrared spectra of the atmospheres of Earth-like planets to be observed. The future ground-based 39-meter Extremely Large Telescope (ELT) will also make it possible to study the composition of atmospheres of extrasolar planets. This telescope is scheduled to be operational in 2030.

## METHODS FOR ANALYZING THE RADIATION SPECTRA OF THE EARTH AND EXO-EARTH

To calculate the radiation spectra of the atmospheres of the Earth and exo-Earh rotating with periods $P = 1$ and 100 days, respectively, we considered the atmospheric circulation lasting for two years. For this, we used the CCM3 (Community Climate System Model version 3; Kiehl et al., 1998) to simulate general circulation in the atmosphere and the SBDART program (Cho and Ipatov, 2008; Ipatov and Cho, 2008, 2013, 2014). We assumed that, in two years, the circulation in the atmosphere of an exoplanet can sufficiently stabilize compared to the initial state. The time interval under consideration is 7.3 times longer than the rotation period of the exoplanet in question. Specifically, the CCM3 takes into account the radiation, water convection, fraction of clouds, types of the terrestrial surface, and temperature and pressure distribution profiles at each point of the planet; and it also calculates the spectrum near the planet's surface. In the CCM3 simulations, we considered 128 longitude values (0°–357°), 64 latitude values (from −88° to 88°), and 18 layers (altitude levels) over the whole surface of the globe. A total of $128 \times 64 = 8192$ cells of size 2.8° × 2.8° were considered on the surface of the planet, and $128 \times 64 \times 18 = 147456$ cells, in the three-dimensional model. The altitude partitioning into 18 layers was not uniform; and the top two layers corresponded to pressures from 83.1425 to 2.917 mbar, i.e., altitudes from 17 km to the level between 50 and 100 km. Details about the altitude partitioning of the layers are discussed by Kiehl et al. (1996, Section 3 and Fig. 1). The time step in the CCM3 program was 20 min. At present, to calculate the atmospheric circulation, there are available fresher programs than the CCM3.

The Santa Barbara DISORT Atmospheric Radiative Transfer (SBDART) code calculates the radiative transfer in the atmosphere with taking into account clouds (Ricchiazzi et al., 1998). In this code, the spectrum for a single point on the Earth's surface (for a pair of longitude and latitude values) is calculated with analyzing the atmosphere at different altitudes. Using the SBDART as a subroutine, we calculated the average spectrum for some region on the planet. We analyzed the spectrum of upward radiation at altitudes $h = 1$ or 11 km (below, we mainly discuss the spectrum at $h = 11$ km) in two wavelength ranges of 0.3–1 and 1–18 μm and for a source planet (the initial state), the Earth, and the exo-Earth. We assumed that the infrared radiation at altitudes of 1 and 11 km fairly well characterizes the radiation near the surface and the radiation leaving the planet, respectively. A typical interval considered in the general circulation models of the Earth's troposphere (i.e., the weather layer) is 1–11 km. Note that, according to Mettler et al. (2023), the radiation characterizing $CH_4$, $CO_2$, and $N_2O$ originates from the atmosphere at pressure levels greater than 200 mbar, i.e., at altitudes not higher than 11 km. The initial conditions for the Earth and exo-Earth were the same. The initial conditions were understood as data about the state of the Earth and its atmosphere (including the temperature and clouds) at some time point. The data were taken from the example CCM3 program, which was used in modeling the atmospheric circulation. The initial and boundary data for this program were discussed by Kiehl et al. (1996, Section 7). The purpose of our studies was to trace changes in the maps of the cloud, temperature, and radiation distributions over the planet, which were induced by its rotation for two years with an axial rotation period of 1 or 100 days.

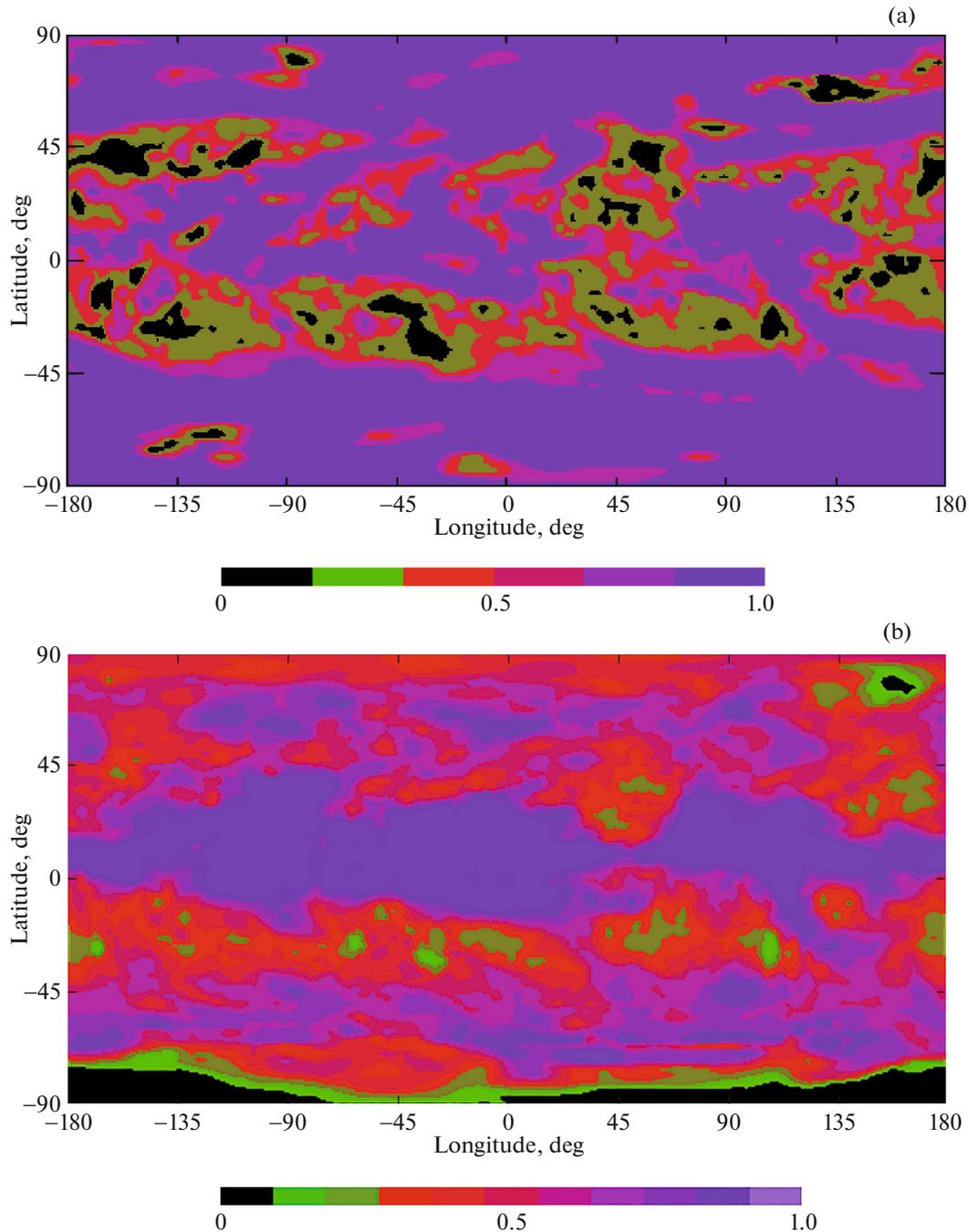

**Fig. 1.** The distribution maps of clouds for longitudes from −180° to 180° and latitudes from −90° to 90° for the models of the Earth ($P = 1$ d) (a) and the exo-Earth ($P = 100$ d) (b). The colors in the bar below the map correspond to the fraction (from 0 to 1) of the planet's surface covered by clouds in one cell of 2.8° × 2.8°.

MODELS FOR ANALYZING THE RADIATION SPECTRA OF THE EARTH AND EXO-EARTH

All calculations of the spectra were performed for the initial state, the Earth, and the exo-Earth. The following models were considered.

**Model 1.** In this model, the average radiation flux depends only on the considered region rather than on the angle, at which this region is observed. For one run of computation with the SBDART code, the weighting factor is proportional to the area of the region (2.8° × 2.8° at the equator) corresponding to the pair ($L$, $B$) of longitude and latitude values. This area is proportional to cos$B$ but independent of $L$. Model 1 corresponds to an abstract case, in which the observer sees all points of the surface at the same angle (perpendicular to the surface of the planet).

**Model 1a.** In this model, we considered all longitude values (0°–360°) and a fixed latitude value. The latitude values were fixed at −88° (the South Pole), −82°, −77°, −66°, −43°, 1° (the equator), 46°, 77°, and 88° (the North Pole).

**Model 1b.** In this model, the average values of the radiation flux were calculated for all of the pairs ($L$, $B$), i.e., the entire surface of the planet was considered.

**Model 1c.** The Southern Hemisphere was considered (−90° ≤ $B$ ≤ 0°, 0° ≤ $L$ ≤ 360°).

**Model 2.** In this model, the average radiation flux depends on the angle at which different parts of the considered region, being half of the planet's surface, are seen. For example, when the region is viewed from the equator at $L = L_0$, the weighting factor $k$ for one SBDART computation run is abs(cos$B$ × cos$B$ × cos($L - L_0$)). In this product, one of the cos$B$ multipliers is due to the fact that the area, corresponding to the ($L$, $B$) pair, is smaller for latitudes closer to the pole (as in Model 1), while the other cos$B$ multiplier is due to the fact that the observer sees areas at different latitudes under different angles. When the region is viewed from the pole, $k$ = abs(cos$B$ × sin$B$).

Model 1 was used to study the radiation flux from different regions of the planet, while Model 2 shows how an observer might see this flux. In the simulations, we studied large-scale hydrostatic three-dimensional global dynamics of the atmosphere that is heated/cooled by radiation.

## DISTRIBUTION MAPS FOR CLOUDS AND TEMPERATURE

Below we consider maps of cloud and temperature distributions over a planet for longitudes from −180° to 180° and latitudes from −90° to 90°. The coverage by clouds was calculated as 1 − [(1 − CLD1) × (1 − CLD2) × ...(1 − CLD18)], where CLD1, ..., CLD18 are values of the cloud coverage at 18 different altitude levels. The cloud distribution maps were different for different rotation periods of the planet (Figs. 1a and 1b). For the Earth there were three large cloud regions (blue color) in the meridian direction, while for the exo-Earth (with $P = 100$ d) there was such a region only near the equator. For this exoplanet, there were almost no clouds near the South Pole, while almost the entire region near the equator was covered with clouds. For the Earth and for the initial state, there were many clouds near both poles. In the cloud distribution map of the Earth, there were fewer clouds near the equator than near the poles. The rotation velocity of the surface of the planet is higher at the equator than at the poles. The rotation velocity of the Earth is 100 times the rotation velocity of the considered exoplanet. Clouds try to leave the regions, where the surface rotates faster, which may explain a smaller number of clouds near the Earth's equator.

In the equatorial region, the maps of the surface temperature (Fig. 2) differed little at $P = 1$ d and $P = 100$ d. We considered the average temperatures over the rotation period of the planet for the near-surface (out of 18) layer of the modeled atmosphere. The values of temperature $T$ were more uniform for the exo-Earth than for the Earth. For example, a region of $T > 290$ K was smaller for the exo-Earth than for the Earth, and a region of $T < 230$ K was absent for the exo-Earth. For the Earth, there were several regions with $T < 230$ K. The main difference in the $T$-maps was near the South Pole. At the South Pole, the temperature was higher for the exo-Earth than for the Earth and the initial state. Near the South Pole, the line $T = 240$ K is rather long for the Earth and is absent for the exo-Earth. There was even a line of $T = 220$ K near the South Pole for the initial state. The differences between the $T$-maps for the exo-Earth and the Earth were smaller than the differences between these maps and the initial $T$-map. The temperature was larger at the North Pole (265 K) than at the South Pole.

## RADIATION SPECTRA OF THE EARTH AND THE EXO-EARTH AT WAVELENGTHS OF 1–18 MICROMETER

Examples of the radiation spectra of the Earth and exo-Earth at wavelengths of 1–18 μm are shown in Figs. 3–5. For the considered axial rotation periods at fixed latitude values in Model 1a (which ignores the orientation of elements of the planet's surface), the radiation flux was maximal at the equator and decreased toward the poles both for the Earth and exo-Earth (Table 1; Figs. 3 and 4). For the source planet in Model 1a, the maximal value of the radiation flux $F_{max}$ (usually achieved at a wavelength of λ ≈ 10 μm) at the South and North Poles was, respectively, 4 and 1.7 times smaller than that at the equator. For the source planet (of both the Earth or the exo-Earth), the maximal radiation flux values $F_{max}$ were almost independent of latitude at latitudes $B$ ≤ −77° (Table 1). For the same planet, the radiation flux values at wavelengths λ ≈ 10–12 μm were 1.1 times greater at latitude $B = -77°$ than at the South Pole, and 1.05 times greater at latitude $B = 77°$ than at the North Pole. For the South Pole on the Earth and exo-Earth, these values exceeded the initial value by a factor of 1.3 and 1.7, respectively. The values of radiation from the both planets differed significantly at wavelengths of about 5–10 and 13–16 μm, e.g., near the poles and at $B$ ≤ −43° and $B$ ≥ 77°. For the North Pole on the Earth, exo-

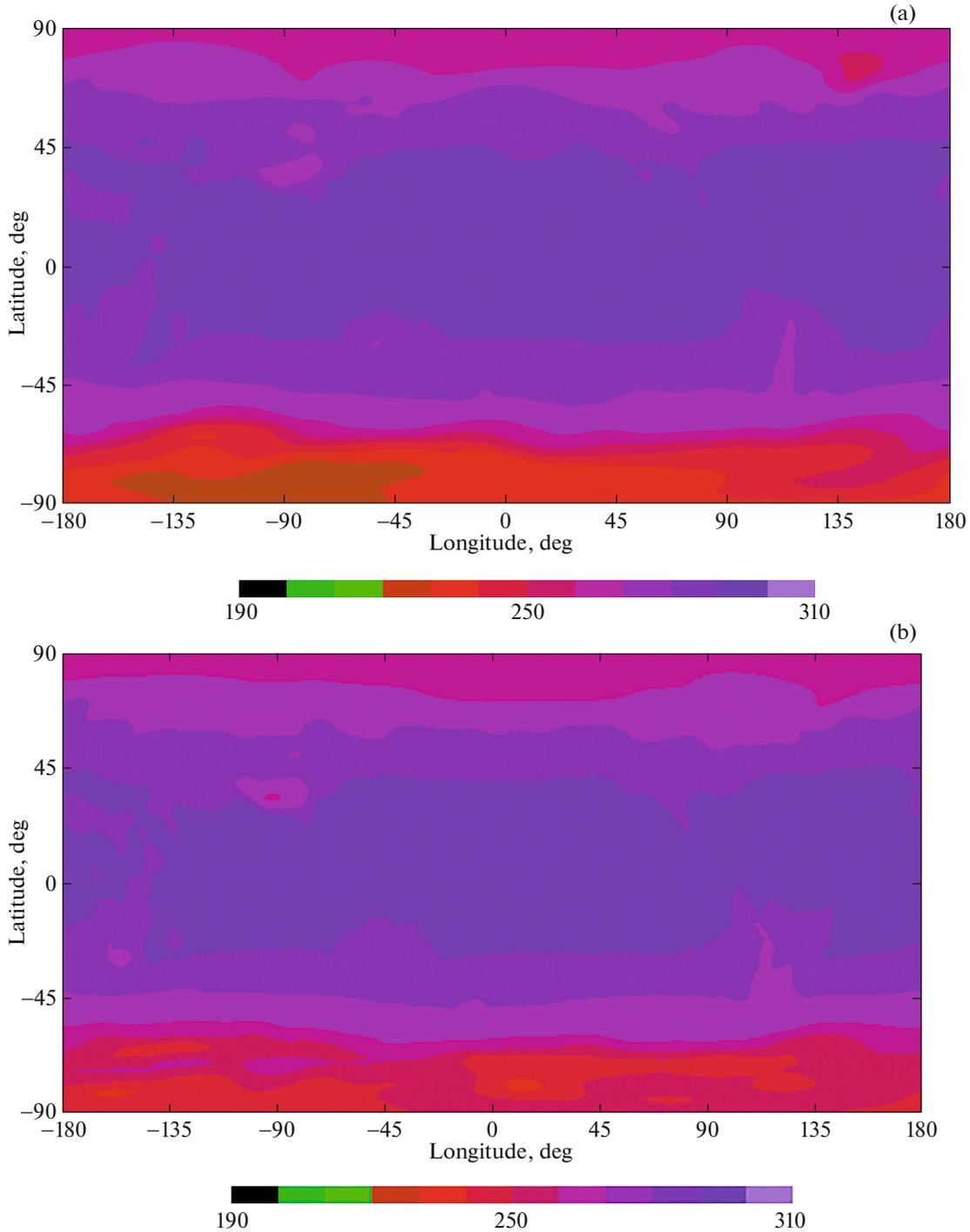

**Fig. 2.** The distribution maps of the surface temperature for longitudes from −180° to 180° and latitudes from −90° to 90° for the models of the Earth ($P = 1$ d) (a) and the exo-Earth ($P = 100$ d) (b). The colors in the bar below the map correspond to the temperature (from 190 to 310 K) of the planet's surface in one cell of 2.8° × 2.8°.

Earth, and source planet, the maximal flux values were close. For the radiation at an altitude of 1 km, the spectra were close for these three planets at −43° ≤ $B$ ≤ 88°.

For two years, the flux for the Earth remained almost unchanged compared to the initial flux (compare Figs. 4a and 4b). The radiation flux at latitude of −77° is much smaller than at the equator (compare Figs. 3 and 4). In an interval of ~5−10 μm, the absorption (the radiation absorption between 1 and 11 km) is smaller in Fig. 4c than in Fig. 4b. If the Earth and the initial state are compared, the differences between the radiation values at altitudes of 11 and 1 km are smaller than those in the case when the exo-Earth and the initial state are compared. In all of the diagrams, for $h = 11$ km, there were a local minimum in the flux at $\lambda \approx 14-16$ μm (absorption by $CO_2$) and a less pronounced local minimum at 9.5 μm (absorption by $O_3$). All of the above features are primarily caused by different coverage by clouds on the exo-Earth and the modeled Earth.

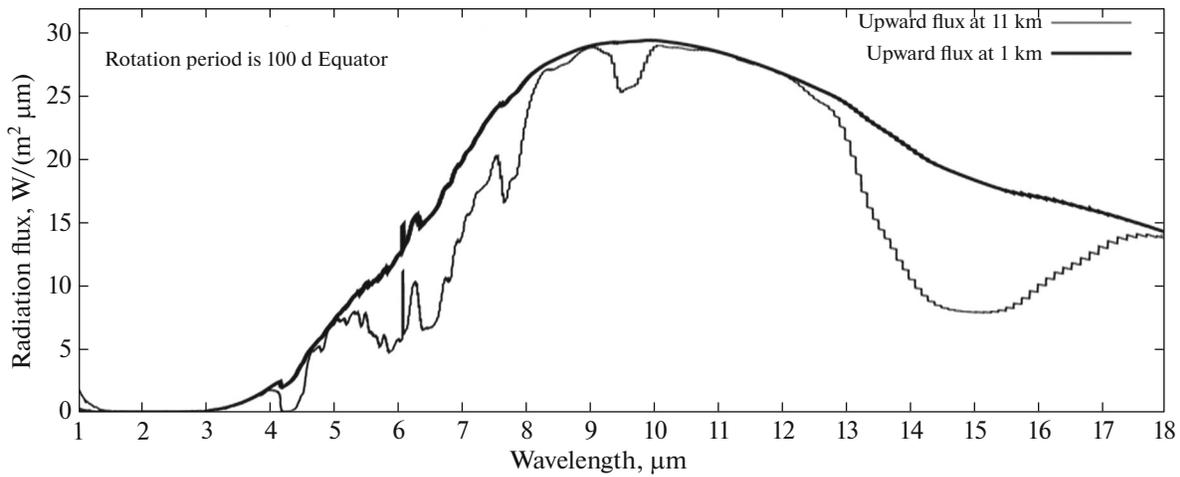

**Fig. 3.** The longitude-averaged spectrum of the upward radiation flux (expressed in W/(m² μm)) at the equator (Model 1a) for the exo-Earth model at wavelengths of 1−18 μm. The spectra are almost identical for the models of the Earth ($P = 1$ d), the exo-Earth ($P = 100$ d), and the initial state. The upper (thicker) curve corresponds to an altitude of 1 km, and the lower curve, to an altitude of 11 km.

For the entire surface of the Earth, exo-Earth, and source planet (Model 1b), the dependences of the radiation flux on $\lambda$ were very close. For the entire surface of the Southern Hemisphere (Model 1c), the $F_{max}$ value was 15 W/(m² μm), which is smaller by 7% than the value for the entire surface (16 W/(m² μm)) (Model 1b). These values are a factor of two lower than that at the equator. This means that there is a difference between the diagrams for the Southern and Northern Hemispheres (e.g., this may be caused by the influence of Antarctica). On the other exoplanets, the topography may be different (there may be no Antarctica analog, but there may be something that is not present on the Earth). The above noted difference between the spectra of the Northern and Southern circumpolar regions of the Earth suggests that, for the other exoplanets, the spectra of this kind may be variable. In Models 1b and 1c, the difference between the $F_{max}$ values for the Earth, exo-Earth, and source planet is smaller (<2%) than that for the two hemispheres. Hence, the effect of the viewing direction on the observed radiation flux may be larger than the effect of the rotation period (see also the results of Model 2). For the Southern Hemisphere, at a wavelength of about 14–16 μm and an altitude of 11 km, the radiation flux for the exo-Earth was 1.2 times larger than for the Earth and the initial state.

**Table 1.** The latitude dependence of the maximal radiation flux at wavelengths of 10−12 μm (Model 1a)

| Latitude, deg | Maximal radiation flux $F_{max}$, W/(m² μm) | | |
|---|---|---|---|
| | initial state | Earth | Exo-earth |
| −88 | 6.5 | 9 | 12 |
| −82 | 8 | 10 | 12.5 |
| −77 | 7.5 | 9.5 | 12.5 |
| −66 | 13 | 13.5 | 14 |
| −43 | 23 | 23 | 22 |
| 1 | 30 | 30 | 29.5 |
| 46 | 25 | 25 | 25 |
| 77 | 18 | 18 | 18 |
| 88 | 17.5 | 17.5 | 17 |

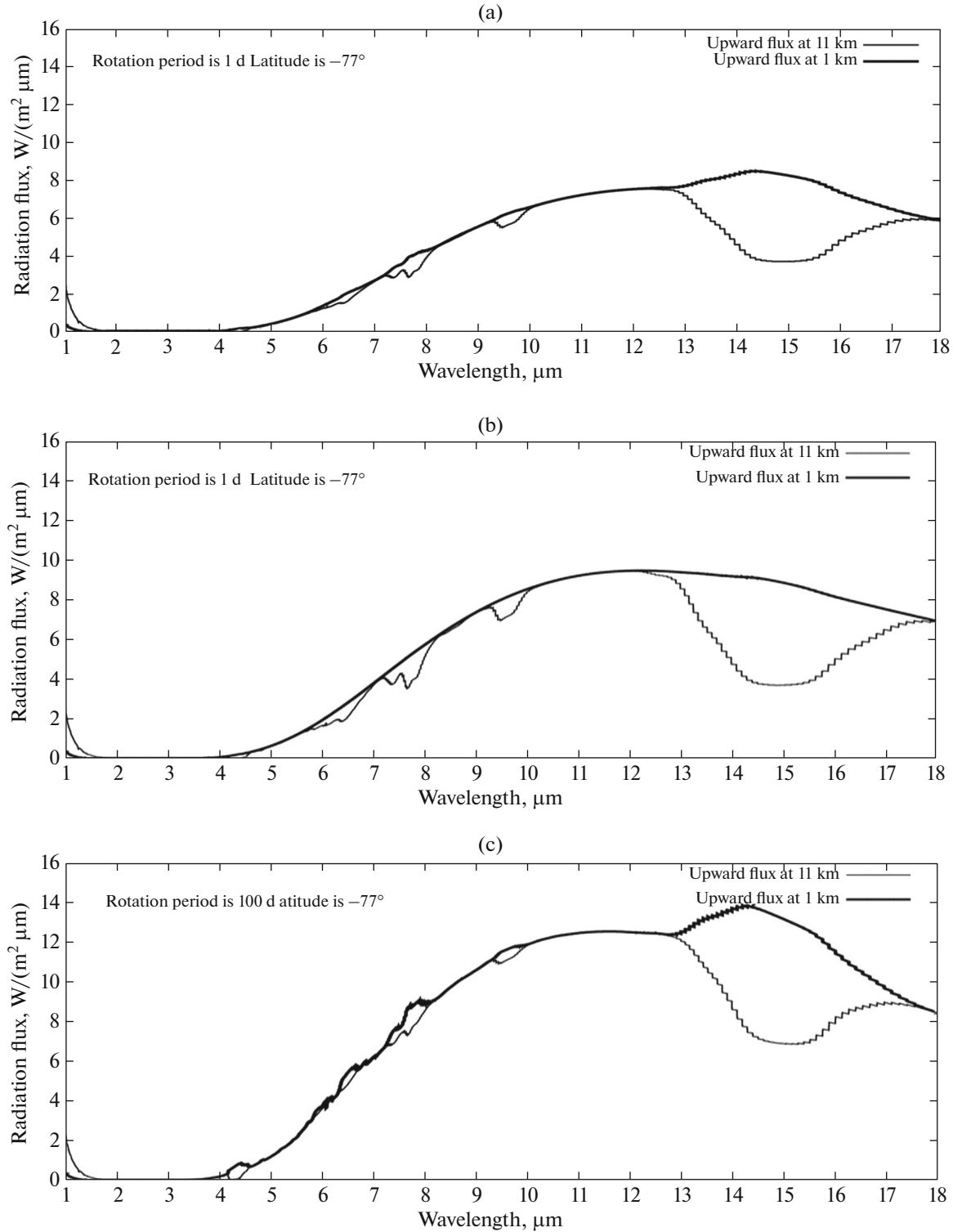

**Fig. 4.** The longitude-averaged spectrum of the upward radiation flux (expressed in W/(m$^2$ μm)) in the 1–18 μm wavelength range for a latitude of −77° (Model 1a) and for the initial state (common for the Earth and exo-Earth) (a), the Earth model ($P = 1$ d) (b), and the exo-Earth model ($P = 100$ d) (c). The upper (thicker) curve corresponds to an altitude of 1 km, and the lower curve, to an altitude of 11 km.

In the spectral diagrams, the maximal value $F_{max}$ of the radiation flux is higher for the surface regions with a higher temperature. For the exo-Earth, the cloud cover was maximal in a wide zone near the equator and was minimum near the South Pole. These particular features of the cloud cover may help to understand why the differences between the radiation fluxes at altitudes of 1 and 11 km for the South Pole differ from those for the equator.

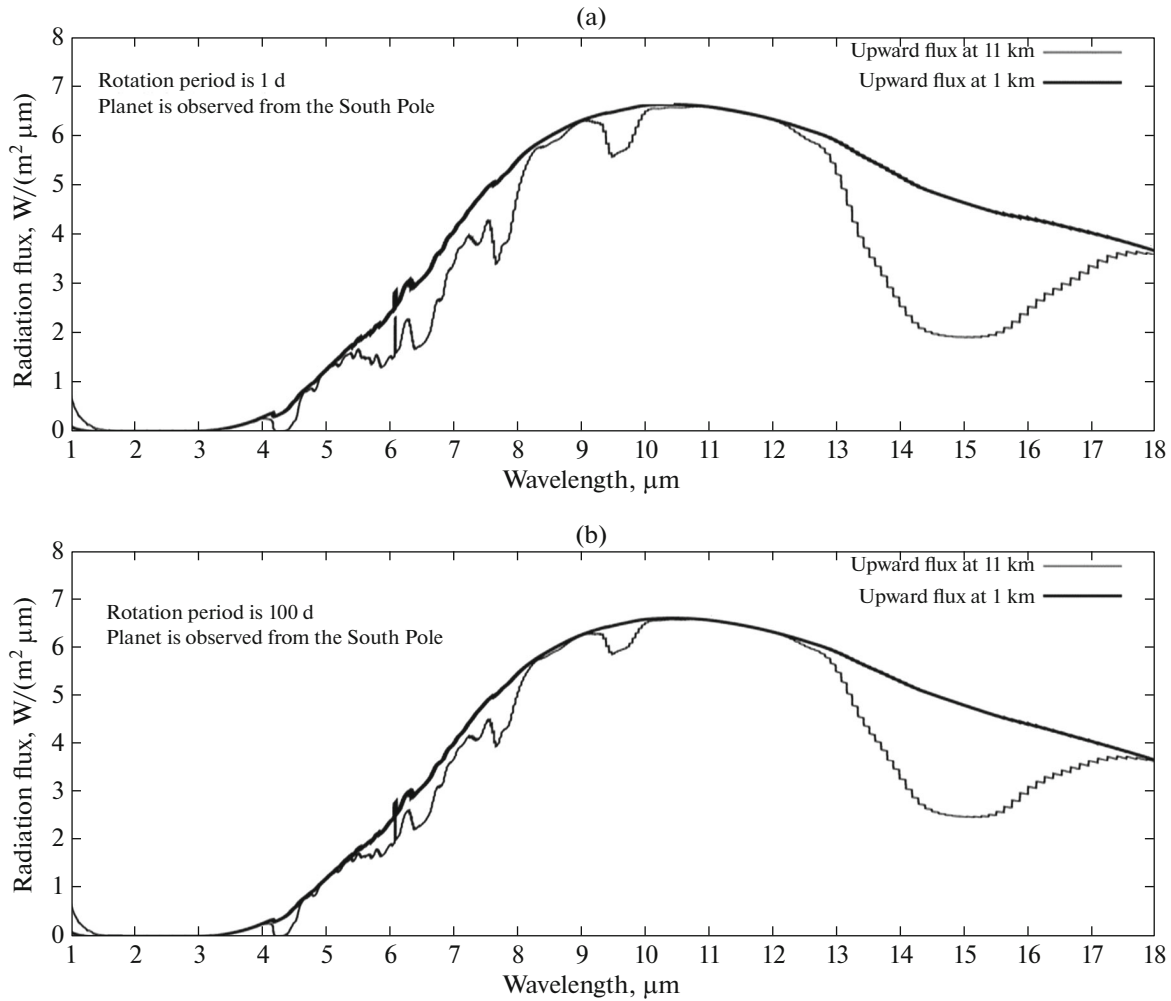

**Fig. 5.** The radiation flux spectrum (expressed in W/(m² μm)) seen from the South Pole (Model 2) in the 1–18 μm wavelength range for the Earth ($P = 1$ d) (a) and the exo-Earth ($P = 100$ d) (b). The upper (thicker) curve corresponds to an altitude of 1 km, and the lower curve, to an altitude of 11 km.

In Model 2, when the planet is observed from the North Pole, $F_{max}$ was 1.2 times larger than for viewing from the South Pole (see Table 2). The difference between the radiation from the Northern and Southern Hemispheres could be due to the influence of Antarctica. The results of Models 1b, 1c, and 2 indicate in favor of the fact that the influence of the viewing direction on the radiation flux observed from the entire planet can be greater than the influence of the rotation period. For the Southern Hemisphere at a wavelength of 14–16 μm and $h = 11$ km, the flux for the exo-Earth was 1.2 times larger than for the Earth and the initial state. The differences in the flux values can be related to the peculiarities in the cloud distribution and to the fact that $F_{max}$ is larger for hotter zones.

**Table 2.** The maximal radiation flux at wavelengths of 10.0–10.4 μm and at an altitude of 1 km (at 11 km, the flux is usually lower by a factor of 1.01–1.02) for different hemispheres of the planet and for several directions of the line of sight to the planet (Model 2)

| Object under study | Viewing direction | | | |
|---|---|---|---|---|
| | south pole | north pole | equator (Latitude $L = 90°$) | Equator (Latitude $L = 270°$) |
| Initial state | 6.61 | 8.09 | 8.27 | 8.49 |
| Earth | 6.63 | 7.97 | 8.05 | 8.31 |
| Exo-Earth | 6.61 | 7.77 | 8.25 | 8.09 |

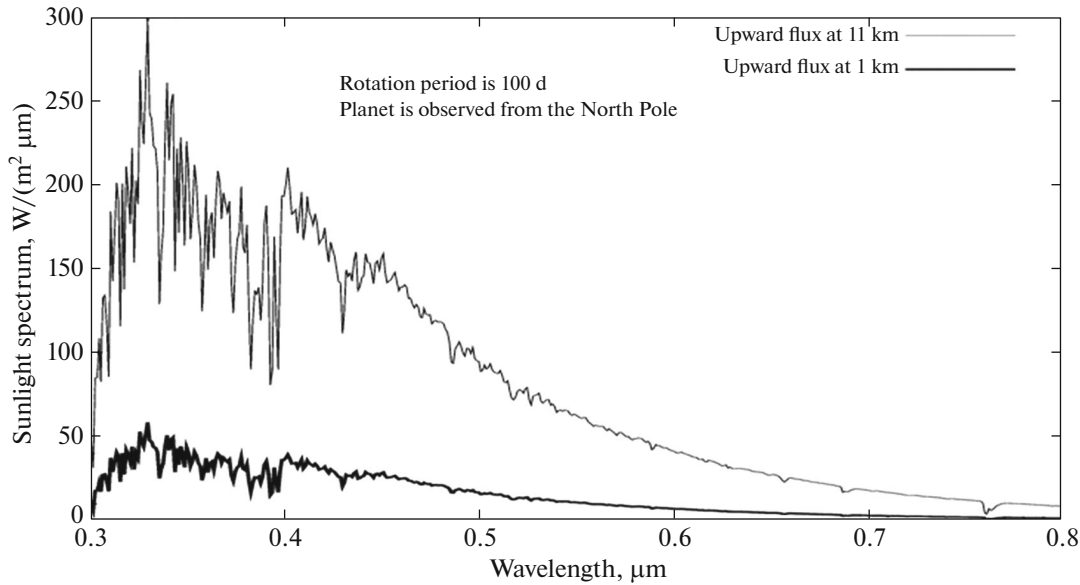

**Fig. 6.** The spectra of visible reflected sunlight (expressed in W/(m$^2$ μm)) in the 0.3–0.8 μm wavelength range for the total upward flux at altitudes of 11 and 1 km, when the exo-Earth is viewed from the North Pole ($P = 100$ d; Model 1a). For the Earth model ($P = 1$ d) and the initial state, the spectra are almost the same as those for the exo-Earth. The upper (thicker) curve corresponds to an altitude of 1 km, and the lower curve, to an altitude of 11 km.

If we take into account the influence of the viewing geometry in observations of the planet (Model 2), the total radiation flux decreases by a factor of two (compared to that in Model 1), since the line of sight is not perpendicular to the surface elements. However, in general, the behavior does not differ from that described above, when the flux for each surface element is simply integrated without correcting for the change in orientation. This is consistent with the similarity of the spectra in the equatorial region on the both planets: the regions at higher latitudes do not contribute significantly. The peak values of the spectra (see Table 2, where all digits are significant) also differ by no more than 5% in the cases, when the observation of the planet is centered on longitude $L = 90°$ or $270°$ and latitude $B = 0°$ (equator) or at the North Pole (with a full range of longitudes). For the view centered on the South Pole, the radiation flux peak is smaller by a factor of 1.2 than that for the view centered on the North Pole, which again indicates the asymmetry in the fluxes from the two poles. When looking at the South Pole, the difference between the radiation fluxes in the ~6–8 μm range at altitudes of 11 and 1 km is smaller for the exo-Earth than for the Earth (Fig. 5). The amplitude of temporal changes in the spectra is smaller, when the exo-Earth is observed from different directions close to the equator than from the other latitudes (excluding those close to the poles). Hence, based on the observations of temporal changes in the spectra of the exo-Earth, we may discuss the viewing angle and/or the rotation period.

For the Earth and the exo-Earth, the following features turned out to be common: (1) the planets have a wide interval of absorption by $CO_2$ around 14 μm; (2) there are no significant differences in the spectra near the equator (however, for some regions, such as those near the poles, there can be significant differences in the spectra); (3) if the spectrum is integrated over the entire planetary disk, the difference in the spectra of the Earth/exo-Earth observed from different directions is much smaller compared to that from observations of individual regions of the planets; however, the difference in the integrated spectral signals from the Earth and the exo-Earth is still noticeable. For example, this difference is noticeable in the spectrum for 11 km observed from the South Pole, but it is small if the entire disk is observed from different equatorial directions. Our results show that the spectral signal cannot be used to determine the rotation velocity of the exoplanet, since its angle of observation is usually not known in advance. For Earth-like exoplanets, the largest differences in the spectra are observed at wavelengths of about 5–10 and 13–16 μm.

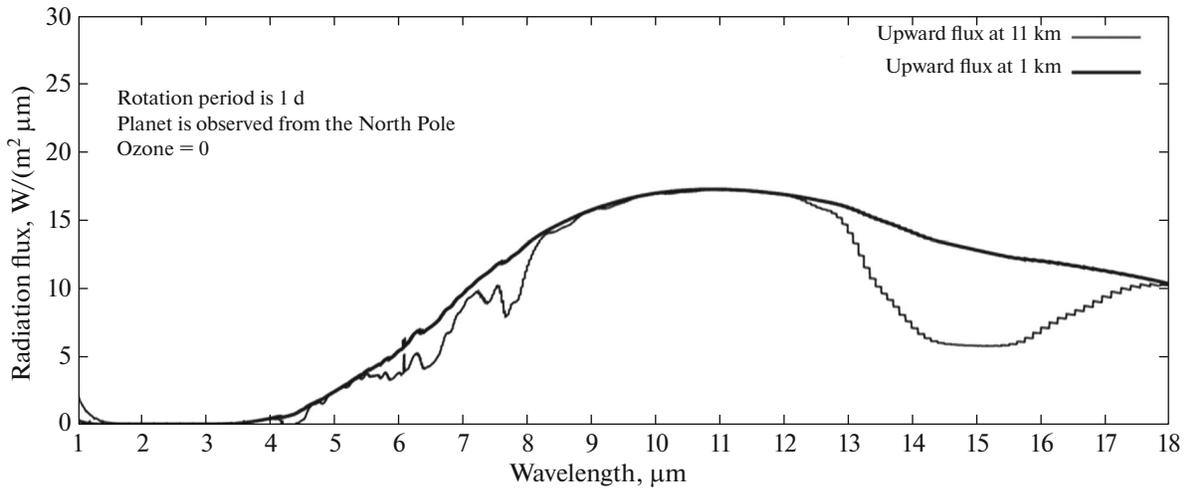

**Fig. 7.** The radiation flux spectrum (expressed in W/(m² μm)) in the 1–18 μm wavelength range for the Earth viewed from the North Pole ($P$ = 1 d; Model 1a); ozone is absent. The upper (thicker) curve corresponds to an altitude of 1 km, and the lower curve, to an altitude of 11 km.

## RADIATION SPECTRA AT WAVELENGTHS SMALLER THAN 1 MICROMETER

In the previous section, we analyzed the diagrams for the wavelengths from 1 to 18 μm (the infrared range). In Fig. 6, we present the spectra of visible reflected sunlight at wavelengths from 0.3 to 0.8 μm for the total upward flux at altitudes of 11 and 1 km. The diagram is shown for the North Pole of the exo-Earth, but the diagrams for the other latitudes (Model 1a) and for the Earth and the initial state look similar to Fig. 6 (for example, all of the diagrams exhibit a local minimum at a wavelength of 0.76 μm). The shape of the spectrum for the 11-km altitude is the same as that for the 1-km altitude, but the values are roughly 5 times larger. It is important to note that in a wavelength interval of 3–18 μm, on the contrary, the flux for 11 km always does not exceed the flux for 1 km. This difference is due to the fact that a part of the sunlight coming to the Earth is reflected from the atmosphere, while a part of the infrared radiation from the Earth's surface is absorbed by the atmosphere.

At wavelengths less than 1 μm, there are almost no differences in the diagrams for the Earth, exo-Earth, and initial state (consequently, these wavelengths are not recommended for studying the differences in the rotation period of planets). In this range, for the equator and the North Pole, the spectral curve for the Earth can lie above the curve for the exo-Earth by no more than 7%; for the South Pole, the difference is smaller (in contrast to the situation at longer wavelengths). The difference between the Earth and the exo-Earth is larger than the difference between the Earth and the initial state. The fluxes may be a few percent smaller for the South Pole than for the North Pole or the equator, while the spectral curves for the equator may lie slightly lower than for the North Pole, which results in a larger relative difference for an altitude of 1 km than for 11 km.

## RADIATION SPECTRA AT DIFFERENT DISTRIBUTIONS OF OZONE

We considered the radiation spectra for a model with some (typical for the Earth) altitude distribution of ozone and for a model, in which the ozone density was assumed to be 0.00005 g/m³ at all altitudes and at all points of the surface. The diagrams for these two models (at wavelengths of 0.3 to 18 μm) were almost identical, and no difference could be seen. This means that, if the ozone distribution is not exactly known, we may consider a model with the ozone density fixed at 0.00005 g/m³.

We also considered a third model, in which the ozone density is zero. The result of this model differed from that of the other two models of the ozone density only in two wavelength intervals. In the diagrams for the total upward radiation flux at an altitude of 11 km, a local minimum at a wavelength of about 9.4–10 μm was absent if the ozone density is zero (Fig. 7), although such a minimum is seen in the diagrams for the other two models (there was no such a difference for the radiation flux at an altitude of 1 km). At wavelengths less than ~0.35 μm (the wavelengths less than 0.25 μm are not considered in the SBDART code), we observe some differences between the diagrams (for the radiation fluxes both at 1 and at 11 km) obtained for the ozone-free model and the other two models. Since ozone is essential for life, the wavelength inter-val around 9.4–10 μm may be important for future observations of Earth-type planets. From the analysis of the spectrum at wavelengths of about 9.4–10 μm, we may conclude whether the atmosphere of the exoplanet contains ozone or not. In this paper, we do not consider in details how the concentration of ozone may influence the radiation from the Earth, since the main ozone layer is located at altitudes higher than 15 km.

## CONCLUSIONS

Using the CCM3 as a general atmospheric circulation model and considering the atmospheric circulation lasting for two years, we calculated the radiation spectra of the Earth and exo-Earth rotating with periods of 1 and 100 days, respectively. In these spectra, the following common features were found: (1) the both planets (the Earth and the exo-Earth having rotation periods of 1 and 100 days, respectively) exhibit a wide absorption band of $CO_2$ centered at ~14 μm; (2) there are no substantial differences in the radiation spectra for near-equatorial regions of the exo-Earth, rotating with a period of 100 days, and the Earth, rotating with a period of 1 day; however, for some regions, (e.g., near the South Pole), the difference may be significant, which suggests that the direction of viewing the planet matters; and (3) if the spectrum is integrated over the entire planetary disk, the differences in the spectral signal obtained from different directions decrease. However, even in the integrated signal, a noticeable difference between the Earth and the exo-Earth can be observed (for example, for an altitude of 11 km, the spectra observed from the South and North Poles are not identical); however, the difference is small, if one observes the whole disk from different directions near the equator. According to our results, the analysis of the radiation spectrum of a planet cannot be used for determining the rotation velocity, since, in general, the geometry of observations is a priori unknown.

Our calculations also show that the radiation bands at about 5–10 and 13–16 μm would be the best wavelength ranges to be considered for observations, at least for the exoplanet model, which is similar to the Earth in all respects except for the rotation period. By analyzing the spectra at wavelengths around 9.5–10 μm, we can infer whether ozone is present in the extrasolar planet or not.

## ACKNOWLEDGMENTS

We are grateful to the reviewers for useful remarks that helped us to improve the paper.

## FUNDING

This study was supported by ongoing institutional funding of the Vernadsky Institute of Geochemistry and Analytical Chemistry of the Russian Academy of Science.

## CONFLICT OF INTEREST

The authors of this work declare that they have no conflicts of interest.